\documentstyle{article}
\input epsf.sty

\def\lsim{\lower0.6ex\vbox{\hbox{$ \buildrel{\textstyle 
<}\over{\sim}\ $}}}
\def\rsim{\lower0.6ex\vbox{\hbox{$ \buildrel{\textstyle 
>}\over{\sim}\ $}}}
\def\hii{H\thinspace{$\scriptstyle{\rm II}$}~}
\def\hi{H\thinspace{$\scriptstyle{\rm I}$}~}
\def\ie{{\it i.e.},~}
\def\eg{{\it e.g.},~}

\def\3he{$^3$He}
\def\4he{$^4$He}
\def\6li{$^6$Li}
\def\7li{$^7$Li}
\def\3h{$^3$H}
\def\Yp{Y$_{\rm P}$~}
\def\etal{{\it et al.~}}

\def\la{\mathrel{\mathpalette\fun <}}
\def\ga{\mathrel{\mathpalette\fun >}}
\def\fun#1#2{\lower3.6pt\vbox{\baselineskip0pt\lineskip.9pt
  \ialign{$\mathsurround=0pt#1\hfil##\hfil$\crcr#2\crcr\sim\crcr}}}
\def\beq{\begin{equation}}
\def\eeq{\end{equation}}

\begin{document}

\begin{center} 

{\Large{\bf Light Element Nucleosynthesis}}
 
\vskip 0.4in
{Gary Steigman}
 
\vskip 0.2in

{\it {Departments of Physics and Astronomy,The Ohio State University, 
\\ Columbus, OH 43210, USA}}\\

\end{center}

\twocolumn

\section{Introduction}

The early Universe was hot and dense behaving as a cosmic nuclear 
reactor during the first twenty minutes of its evolution.  It was,
however, a ``defective" nuclear reactor, expanding and cooling very 
rapidly.  As a result, only a handful of the lightest nuclides were 
synthesized before the density and temperature dropped too low for 
the nuclear reaction rates to compete with the universal expansion 
rate.  After hydrogen ($^1$H $\equiv$ protons) the next most abundant 
element to emerge from the Big Bang is helium (\4he $\equiv$ alpha 
particles).  Isotopes of these nuclides (deuterium and helium-3) are 
the next most abundant primordially.  Then there is a large gap to 
the much lower abundance of lithium-7.  The relative abundances of 
all other primordially-produced nuclei are very low, much smaller 
than their locally observed (or, presently observable!) abundances.  
After a brief description of the early evolution of the Universe 
emphasizing those aspects most relevant to primordial, or ``Big Bang" 
nucleosynthesis (BBN), the predicted abundances of the light nuclides 
will be presented as a function of the one ``free" parameter (in 
the simplest, ``standard" model: SBBN), the nucleon (or ``baryon") 
abundance.  Then, each element will be considered in turn in a 
confrontation between the predictions of SBBN and the observational 
data.  At present (Summer 1999) there is remarkable agreement between 
the SBBN predictions of the abundances of four nuclides (D, $^3$He, 
$^4$He, and $^7$Li) and their primordial abundances inferred from 
the observations.  However, there are some hints that this concordance 
of the hot big bang model may be imperfect, so we will also explore 
some variations on the theme of the standard model with regard to 
their modifications of the predicted primordial abundances of the 
light elements.

In the simplest, standard, hot big bang model the currently observed 
large-scale isotropy and homogeneity of the Universe is assumed to 
apply during earlier epochs in its evolution.  Given the currently 
observed universal expansion and the matter and radiation (CBR: 
``cosmic background radiation", the 2.7K ``black body radiation") 
content, it is a straightforward application of classical physics 
to extrapolate back to earlier epochs in the history of the Universe.  
At a time of order 0.1 s after the expansion began the Universe was 
filled with a hot, dense plasma of particles.  The most abundant were 
photons, electron-positron pairs, particle-antiparticle pairs of all 
known ``flavors" of neutrinos ($\nu_{e}$, $\nu_{\mu}$, and $\nu_{\tau}$), 
and trace amounts of neutrons and protons (``nucleons" or ``baryons").  
At such early times the thermal energy of these particles was very high, 
of order a few MeV.  With the exception of the nucleons, it is known or
assumed that all the other particles present were extremely relativistic
at this time.  Given their high energies (and velocities close to, or
exactly equal to the speed of light) and high densities, the electroweak
interactions among these particles were sufficiently rapid to have
established thermal equilibrium.  As a result, the numbers and
distributions (of momentum and energy) of all these particles is
accurately predicted by well-known physics.

\section{Nucleosynthesis In The Early Universe}

The primordial yields of light elements are determined by the competition 
between the expansion rate of the Universe (the Hubble parameter $H$) and
the rates of the weak and nuclear reactions.  It is the weak interaction,
interconverting neutrons and protons, that largely determines the amount
of \4he which may be synthesized, while detailed nuclear reaction rates
regulate the production (and destruction) of the other light elements.
In the standard model of cosmology the early expansion rate is fixed by
the total energy density $\rho$,
\beq
H^2 = 8\pi G \rho/3,
\label {H}
\eeq
where $G$ is Newton's gravitational constant.  In the standard 
model of particle physics the early energy density is dominated 
by the lightest, relativistic particles.  For the epoch when 
the Universe is a few tenths of a second old and older, and 
the temperature is less than a few MeV,
\beq
\rho = \rho_\gamma + \rho_e + {\rm N}_\nu \rho_\nu,
\label {rho}
\eeq
where $\rho_\gamma$, $\rho_e$, and $\rho_\nu$ are the energy densities
in photons, electrons and positrons, and massless neutrinos and antineutrinos 
(one species), respectively; $N_\nu$ is the number of massless (or, very 
light: $m_{\nu} \ll 1$~MeV) neutrino species which, in standard BBN, is 
exactly 3.  In considering variations on the theme of the standard model,
it is useful to allow $N_\nu$ to differ from 3 to account for the presence
of ``new" particles and/or any suppression of the standard particles (\eg
if the $\tau$ neutrino should have a large mass).  Since the energy density
in relativistic particles scales as the fourth power of the temperature,
the early expansion rate scales as the square of the temperature with a
coefficient that depends on the number of different relativistic species.
The more such species, the faster the Universe expands, the earlier (higher
temperature) will the weak and nuclear reactions drop out of equilibrium.
It is useful to write the total energy density in terms of the photon
energy density and $g$, the equivalent number of relativistic degrees 
of freedom (\ie helicity states, modulo the different contributions to 
the energy density from fermions and bosons),
\beq
\rho \equiv (g/2)\rho_{\gamma}.
\label {g}
\eeq
In the standard model at $T \sim 1$ MeV, $g_{SM} = 43/4$.
Account may be taken of additional degrees of freedom by comparing
their contribution to $\rho$ to that of one additional light neutrino
species.
\beq
\Delta \rho \equiv \rho_{TOT} - \rho_{SM} \equiv \Delta N_{\nu}\rho_{\nu}.
\label {Nnu}
\eeq
If the early energy density deviates from that of the standard model, the
early expansion rate (or, equivalently, the age at a fixed temperature) 
will change as well.  The ``speed-up" factor $\xi \equiv H/H_{SM}$ may
be related to $\Delta N_{\nu}$ by,
\beq
\xi = (\rho/\rho_{SM})^{1/2} = (1 + 7\Delta N_{\nu}/43)^{1/2}.
\label {xi}
\eeq


As we'll see shortly, the \4he abundance is very sensitive to the early
expansion rate while the abundances of the other light nuclides depend
mainly on the nuclear reaction rates which scale with the nucleon (baryon)
density.  Since the baryon density is always changing as the Universe
expands, it is convenient to distinguish between models with different
baryon densities using a dimensionless parameter which is either conserved 
or, changes in a known and calculable fashion.  From the very early Universe 
till now the number of baryons in a comoving volume has been preserved and 
the same is roughly true for photons since the end of BBN.  Therefore, the 
ratio of number densities of baryons ($n_{\rm B}$) and photons ($n_{\gamma}$) 
provides just such a measure of the universal baryon abundance.
\beq
\eta \equiv (n_{\rm B}/n_{\gamma})_{0} \ ; \ \eta_{10} \equiv 10^{10}\eta.
\label {eta}
\eeq
The universal density of photons at present (throughout this article the
present epoch is indicated by the subscript `0') is dominated by those in 
the CBR (for T$_{0} = 2.73$~K, $n_{\gamma 0} = 412~$cm$^{-3}$) so that the 
baryon density parameter $\Omega_{\rm B} \equiv (\rho_{\rm B}/\rho_{c})_{0}$, 
the ratio of the present baryon density ($\rho_{\rm B}$) to the present 
critical density ($\rho_{c}$), may be related to $\eta$ and the present 
value of the Hubble parameter H$_{0} \equiv 100h~$kms$^{-1}$Mpc$^{-1}$,
\beq
\eta_{10} = 273\Omega_{\rm B}h^2.
\label {omegab}
\eeq
It should be noted that prior to electron-positron annihilation there 
were fewer photons in every comoving volume (by a factor very close to 
4/11); this is automatically accounted for in all numerical BBN codes.  
It is simply a matter of consensus and convenience that the baryon 
abundance is quoted in terms of its present value.

In SBBN (\ie N$_{\nu} = 3$) the abundances of the light nuclides 
synthesized primordially depend on only one ``free" parameter, $\eta$.
SBBN is thus ``overconstrained" since one value (or, a narrow range 
of values set by the observational and theoretical reaction rate 
uncertainties) of $\eta$ must account consistently for the primordial 
abundances of D, \3he, \4he and \7li.  At the same time this value/range 
of $\eta$ must be consistent with current estimates of (or, bounds to) 
the present baryon density.  For these reasons BBN is one of the key 
pillars supporting the edifice of the standard model of cosmology and, 
it is the only one which offers a glimpse of the earliest evolution of 
the Universe.  In the following we'll first identify the key landmarks 
in the first 20 minutes in the evolution of the Universe in order to 
identify the physical processes responsible for determining the primordial 
abundances of the light nuclides.  Then, after presenting the SBBN 
predictions (as a function of $\eta$; see Fig.~1) we will review the 
current status of the observational data, as well as the steps necessary 
in order to go from ``here and now" to ``there and then" when using 
the data to infer the true primordial abundances.  Then we will be 
in a position to assess the consistency of the standard model.

\begin{figure}[h]
	\centering
	\epsfysize=4.9truein 
\hskip .5in 
\epsfbox{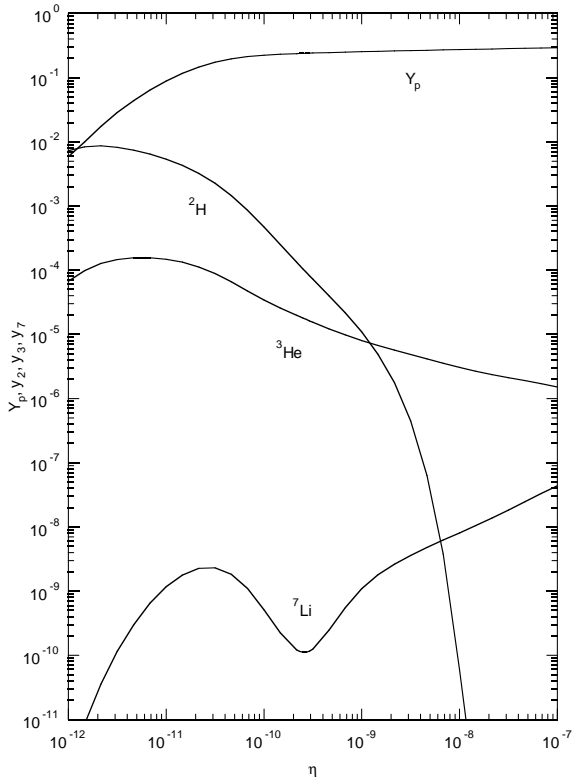}
	\caption{The predicted primordial abundances as a function of $\eta$.
	\Yp is the \4he mass fraction while $y_{2\rm P}$, $y_{3\rm P}$, 
	$y_{7\rm P}$ are the number density ratios to hydrogen of D, \3he,
	and \4he respectively.}
	\label{fig1}
\end{figure}

\subsection {Weak Equilibrium and the \4he Abundance}

Consider now those early epochs when the Universe was only a few tenths 
of a second old and the radiation filling it was at a temperature (thermal 
energy) of a few MeV.  According to the standard model, at those early 
times the Universe was a hot, dense ``soup" of relativistic particles 
(photons, e$^{\pm}$ pairs, 3 ``flavors" (e, $\mu$, $\tau$) of 
neutrino-antineutrino pairs) along with a trace amount (at the level 
of a few parts in $10^{10}$) of neutrons and protons.  At such high 
temperatures and densities both the weak and nuclear reaction rates 
are sufficiently rapid (compared to the early Universe expansion rate) 
that all particles have come to equilibrium.  A key consequence of 
equilibrium is that the earlier history of the evolution of the Universe 
is irrelevant for an understanding of BBN.  When the temperature drops 
below a few MeV the weakly interacting neutrinos effectively decouple 
from the photons and e$^{\pm}$ pairs, but they still play an important 
role in regulating the neutron-to-proton ratio.  

At high temperatures, neutrons and protons are continuously interconverting 
via the weak interactions: $n + e^+  \leftrightarrow  p + {\bar \nu_e},~ 
n + \nu_e  \leftrightarrow  p + e^- ,$ and $n  \leftrightarrow  p + e^- + 
{\bar \nu_e}.$  When the interconversion rate is faster than the expansion 
rate, the neutron-to-proton ratio tracks its equilibrium value, decreasing 
exponentially with temperature ($n/p = e^{-\Delta m/T}$, where $\Delta m = 
1.29$~MeV is the neutron-proton mass difference).  A comparison of the weak
rates with the universal expansion rate reveals that equilibrium may be 
maintained until the temperature drops below $\sim 0.8$~MeV.  When the 
interconversion rate becomes less than the expansion rate, the $n/p$ ratio
effectively ``freezes-out'' (at a value of $\approx 1/6$), thereafter 
decreasing slowly, mainly due to free neutron decay. 

Although $n/p$ freeze-out occurs at a temperature below the deuterium
binding energy, $E_B = 2.2$ MeV, the first link in the nucleosynthetic 
chain, $p + n \rightarrow$ D $+~\gamma$, is ineffective in jump-starting 
BBN since the photodestruction rate of deuterium ($\propto n_\gamma 
e^{-E_B/T}$) is much larger than the deuterium production rate ($\propto 
n_B$) due to the very large universal photon-to-baryon ratio ($\ga 10^9$).  
Thus, the Universe must ``wait" until there are so few sufficiently
energetic photons that deuterium becomes effectively stable against 
photodissociation.  This occurs for temperatures $\lsim 80$ keV, at which 
time neutrons are rapidly incorporated into \4he with an efficiency of 
99.99\%.  This efficiency is driven by the tight binding of the \4he 
nucleus, along with the roadblock to further nucleosynthesis imposed 
by the absence of a stable nucleus at mass-5.  By this time (T $\lsim 
80$ keV), the $n/p$ ratio has dropped to $\sim 1/7$, and simple counting 
(2 neutrons in every \4he nucleus) yields an estimated primordial \4he 
mass fraction
\beq
Y_{\rm P} \approx {2(n/p) \over \left[ 1 + (n/p) \right]} \approx 0.25.
\label{ynp}
\eeq 
As a result of its large binding energy and the gap at mass-5, the 
primordial abundance of \4he is relatively insensitive to the nuclear 
reaction rates and, therefore, to the baryon abundance ($\eta$).  As 
may be seen in Figure 1, while $\eta$ varies by orders of magnitude, 
the predicted \4he mass fraction, \Yp, changes by factors of only a few.  
Indeed, for $1 \leq \eta_{10} \leq 10$, 0.22 $\leq Y_{\rm P} \leq 0.25$.
As may be seen in Figures~1 and 2, there is a very slight increase in \Yp
with $\eta$.  This is mainly due to BBN beginning earlier, when there are
more neutrons available to form \4he, if the baryon-to-photon ratio is
higher.  The increase in \Yp with $\eta$ is logarithmic; over most of the
interesting range in $\eta$, $\Delta$\Yp $\approx 0.01\Delta\eta/\eta$.

\begin{figure}[ht]
	\centering
	\epsfysize=2.4truein 
\epsfbox{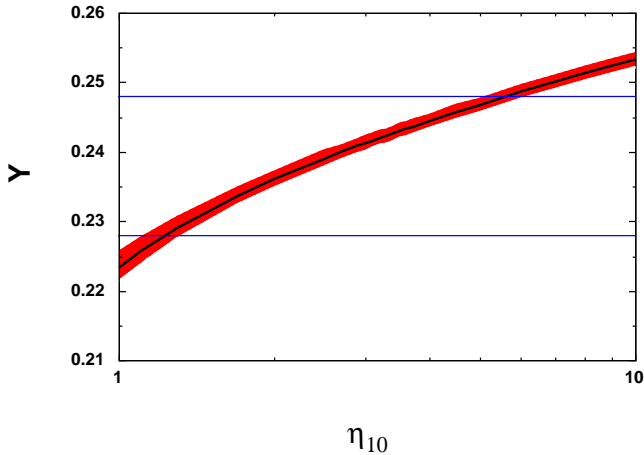}
	\caption{The predicted \4he abundance (solid curve) and the $2 \sigma$
theoretical uncertainty \protect\cite{hata}. The horizontal lines show the
range indicated by the observational data.}
	\label{fig2}
\end{figure}

The \4he abundance is, however, sensitive to the competition between 
the universal expansion rate ($H$) and the weak interaction rate 
(interconverting neutrons and protons).  If the early Universe should 
expand faster than predicted for the standard model, the weak interactions 
will drop out of equilibrium earlier, at a higher temperature, when the 
$n/p$ ratio is higher.  In this case, more neutrons will be available to 
be incorporated into \4he and \Yp will increase.  Numerical calculations 
show that for a modest speed-up ($\Delta N_{\nu} \lsim 1$), $\Delta$\Yp 
$\approx 0.013\Delta N_{\nu}$.  Hence, constraints on \Yp (and $\eta$) 
lead directly to bounds on $\Delta N_{\nu}$ and, on particle physics 
beyond the standard model ~\cite{ssg}.  

It should be noted that the uncertainty in the BBN-predicted mass 
fraction of \4he is very small and almost entirely dominated by the 
(small) uncertainty in the $n-p$ interconversion rates.  These rates 
may be ``normalized" through the neutron lifetime, $\tau_{n}$, whose 
current standard value is 887$\pm$2~s (actually, 886.7$\pm$1.9~s).  
To very good accuracy, a 1~s uncertainty in $\tau_{n}$ corresponds 
to an uncertainty in \Yp of order $2 \times 10^{-4}$.  At this tiny
level of uncertainty it is important to include finite mass, zero- and 
finite-temperature radiative corrections, and Coulomb corrections to 
the weak rates.  However, within the last few years it emerged that 
the largest error in the BBN-prediction of \Yp was due to a too large
time-step in the numerical code.  With this now under control, it is
estimated that the residual theoretical uncertainty (in addition to
that from the uncertainty in $\tau_{n}$) is of the order of 2 parts 
in $10^4$.  Indeed, a comparison of two major, independent BBN codes 
reveals agreement in the predicted values of \Yp to 0.0001 $\pm$ 0.0001 
over the entire range $1 \leq \eta_{10} \leq 10$.  In Figure~2 is 
shown the BBN-predicted \4he mass fraction, \Yp, as a function of 
$\eta$; the thickness of the band is the $\pm 2\sigma$ theoretical 
uncertainty.  For $\eta_{10} \ga 2$ the 1$\sigma$ theoretical uncertainty
in \Yp is $\lsim 6 \times 10^{-4}$.  As we will soon see, the current 
observational uncertainties in \Yp are much larger (see, also, Fig.~2).

\subsection{Deuterium -- The Ideal Baryometer} 
 
As may be seen in Fig.~1, the deuterium abundance (the ratio, by number, 
of deuterium to hydrogen: hereinafter, (D/H)$_{\rm P} \equiv y_{2\rm P}$)
is a monotonic, rapidly decreasing function of the baryon abundance
$\eta$.  The reason for this behavior is easily understood.  Once BBN
begins in earnest, when the temperature drops below $\sim 80$~keV, D
is rapidly burned to \3h, \3he and \4he.  The higher the baryon abundance, 
the faster the burning and the less D survives.  For $\eta_{10}$ in the 
``interesting" range 1 -- 10,  $y_{2\rm P}$ decreases with the $\sim$ 1.6 
power of $\eta$.  As a result, a 10\% error in $y_{2\rm P}$ corresponds 
to only a 6\% error in $\eta$.  This strong dependence of $y_{2\rm P}$ 
on $\eta_{10}$, combined with the simplicity of the evolution of D/H in 
the epochs following BBN, is responsible for the unique role of deuterium 
as a baryometer~\cite{els}.  Because almost all the relevant reaction 
cross sections are measured in the laboratory at energies comparable to 
those of BBN, the theoretical uncertainties in the BBN-predicted abundance 
of deuterium is quite small, 8 - 10\% for most of the interesting $\eta$ 
range shown in Fig.~3.

\begin{figure}[ht]
	\centering
	\epsfysize=2.4truein 
\epsfbox{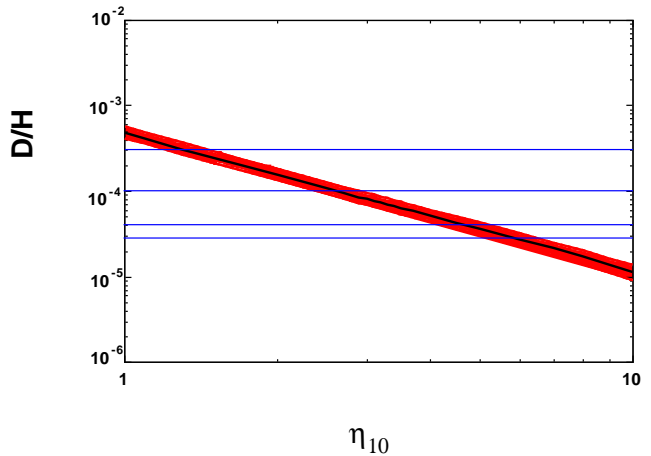}
	\caption{The predicted D/H abundance (solid curve) and the $2 \sigma$
theoretical uncertainty \protect\cite{hata}. The horizontal lines show the
range indicated by the observational data for both the high D/H (upper two
lines ) and low D/H (lower two lines).}
	\label{fig3}
\end{figure}

Deuterium and helium-4 are complementary, forming the crucial link 
in testing the consistency of BBN in the standard model.  While the 
primordial-D abundance is very sensitive to the baryon density, the 
primordial-\4he abundance is relatively insensitive to $\eta$.  
Deuterium provides a bound on the universal baryon density while 
helium-4 constrains the early expansion rate of the Universe, 
offerring bounds on particle physics beyond the standard model.

\subsection{Helium-3 -- Complicated Evolution}

As may be seen in Fig.~1, the predicted primordial abundance of \3he 
behaves similarly to that of D, decreasing monotonically with $\eta$.
Again, the reason is the same: \3he is being burned to the more tightly
bound nucleus \4he.  The higher the nucleon abundance, the faster the
burning and the less \3he survives.  In contrast to the behavior of
D/H versus $\eta$, the decrease of \3he/H with increasing $\eta$ is 
much slower.  This is simply a reflection of the much tighter binding 
of \3he compared to D.  Although the BBN predictions for the abundance
of \3he have similarly small uncertainties (8 - 10\%) to those of D,
it is much more difficult to exploit the observations of \3he to test
and constrain BBN.  The complication is the evolutionary history of
the \3he abundance since BBN.

Although any deuterium cycled through stars is burned to \3he during
the stars' pre-main sequence evolution, \3he will survive in the cooler
stellar exteriors while being destroyed in the hotter interiors.  For
the more abundant lower mass stars which are cooler, a larger fraction
of prestellar \3he (along with the \3he produced from prestellar D) 
survives.  Indeed, for sufficiently low-mass stars (less than a few 
solar masses) incomplete burning actually leads to a buildup of newly
synthesized \3he (to be contrasted with the prestellar D and \3he) which
may -- or may not -- be returned to the interstellar medium.  In fact,
some planetary nebulae are observed to be highly enriched in \3he.
So, the evolution of \3he is complex with stellar destruction competing
with primordial and stellar production.  Indeed, if all low mass stars
were as prolific producers of \3he as indicated by some planetary
nebulae, the solar system and local interstellar medium abundances of
\3he should far exceed those inferred from observations.  Thus, at 
least some low mass stars must be net destroyers of \3he.  Given this
necessarily complex and uncertain picture of production, destruction 
and survival, it is difficult to use current observational data to
infer the primordial abundance of \3he.  Unless and until \3he is
observed in high redshift (\ie early Universe), low metallicity 
(\ie nearly unevolved) systems, it will provide only a weak check 
on the consistency of BBN.  

\subsection{Lithium-7 -- The Lithium Valley}

The trend of the BBN-predicted primordial abundance of lithium
(almost entirely \7li) with $\eta$ is more `interesting' than
that of the other light nuclides (see Figs.~1 \& 4).  The ``lithium
valley", centered near $\eta_{10} \approx 2 - 3$, is the result of
the competition between production and destruction in the two paths
to mass-7 synthesis.  At relatively low baryon abundance ($\eta_{10}
\lsim 2$) mass-7 is mainly synthesized as \7li via $^3$H $+ ^4$He 
$\rightarrow ^7$Li + $\gamma$.  As the baryon abundance increases at
low $\eta$, \7li is destroyed rapidly by $(p,\gamma)2\alpha$ reactions.
Hence the decrease in (Li/H)$_{\rm P} \equiv y_{7\rm P}$ with increasing
$\eta$ seen (at low $\eta$) in Figs.~1 \& 4.  Were this the only route
to primordial synthesis of mass-7, this monotonic trend would continue,
similar to those for D and \3he.  However, mass-7 may also be synthesized
via $^3$He $+ ^4$He $\rightarrow ^7$Be + $\gamma$.  The $^7$Be will later
capture an electron to become \7li.  This channel is very important 
because it is much {\it easier} to destroy \7li than $^7$Be.  As a result,
for relatively high baryon abundance ($\eta_{10} \ga 3$) this latter
channel dominates mass-7 production and $y_{7\rm P}$ increases with
increasing $\eta$.

\begin{figure}[ht]
	\centering
	\epsfysize=2.4truein 
\epsfbox{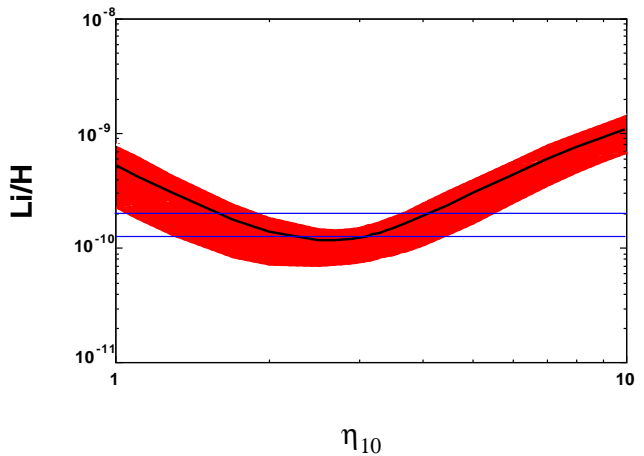}
	\caption{The predicted \7li abundance (solid curve) and the $2 \sigma$
theoretical uncertainty \protect\cite{hata}. The horizontal lines show the
range indicated by the observational data.}
	\label{fig4}
\end{figure}

As may been seen in Figure 4, the BBN-predicted uncertainties for 
$y_{7\rm P}$ are much larger than those for D, \3he, or \4he.  In 
the interval $1 \leq \eta_{10} \leq 10$ the 1$\sigma$ uncertainties 
are typically $\sim 20\%$, although in a narrow range of $\eta$ near 
the bottom of the `valley' they are somewhat smaller ($\sim 12 - 15\%$).

\section{From Here And Now To There And Then}

To test the consistency of SBBN requires that we confront the 
predictions with the primordial abundances of the light nuclides 
which, however, are not {\it observed} but, rather, are {\it 
inferred} from observations.  The path from the observational data 
to the primordial abundances is long and twisted and often fraught 
with peril.  In addition to the usual statistical and systematic 
uncertainties, it is crucial to forge a connection from ``here and 
now" to ``there and then"; \ie to relate the derived abundances to 
their primordial values.  It is indeed fortunate that each of the 
key elements is observed in different astrophysical sites using very 
different astronomical techniques.  Also, the corrections for chemical 
evolution differ among them and, even more important, they can be 
minimized.  For example, deuterium (and hydrogen) is mainly observed 
in cool, neutral gas (so called \hi regions) via UV absorption from 
the atomic ground state (the Lyman series), while radio telescopes 
allow helium-3 to be studied utilizing the analog of the hydrogen 
21~cm line for singly-ionized $^3$He in regions of hot, ionized gas 
(so called \hii regions).  Helium-4 is probed using the emission from 
its optical recombination lines formed in \hii regions.  In contrast, 
lithium is observed in the absorption spectra of warm, low-mass halo 
stars.  With such different sites, with the mix of absorption/emission, 
and with the variety of telescopes and different detectors involved, 
the possibility of correlated errors biasing the comparison with the 
predictions of BBN is unlikely.  This favorable situation extends to 
the obligatory evolutionary corrections.  For example, although until 
recently observations of deuterium were limited to the solar system 
and the Galaxy, mandating uncertain evolutionary corrections to infer 
the pregalactic abundance, the Keck and Hubble Space telescopes have 
begun to open the window to deuterium in high-redshift, low-metallicity, 
nearly primordial regions (Lyman-$\alpha$ clouds).  Observations of 
$^4$He in chemically unevolved, low-metallicity ($\sim 1/50$ of solar) 
extragalactic \hii regions permit the evolutionary correction to be 
reduced to the level of the statistical uncertainties.  The abundances 
of lithium inferred from observations of the very metal-poor halo stars 
(one-thousandth of solar abundance and even lower) require almost no 
correction for chemical evolution.  On the other hand, as noted earlier, 
the status of helium-3 is in contrast to that of the other light elements.  
For this reason, \3he will not be used quantitatively in this article.

The currently very favorable observational and evolutionary situation for 
the nuclides produced during BBN is counterbalanced by the likely presence 
of systematic errors in the path from observations to primordial abundances.  
By their very nature, such errors are difficult -- if not impossible -- to 
quantify.  In the key case of deuterium there is a very limited set of 
the most useful data.  As a result, and although cosmological abundance
determinations have taken their place in the current ``precision" era of
cosmology, it is far from clear that the present abundance determinations 
are truly ``accurate".  Thus, the usual {\it caveat emptor} applies to any 
conclusions drawn from the subsequent comparison between the predictions 
and the data.  With this caution in mind the current status of the data 
will be surveyed in order to infer ``reasonable" ranges for the primordial 
abundances of the key light elements.

\subsection{Deuterium}

Since deuterium is completely burned away whenever it is cycled through 
stars, and there are no astrophysical sites capable of producing deuterium 
in anywhere near its observed abundance \cite{els}, any D-abundance 
derived from observational data provides a {\it lower} bound to its 
primordial abundance. Thus, without having to correct for Galactic 
evolution, the `here-and-now' deuterium abundance inferred from UV 
observations along 12 lines-of-sight in the local interstellar medium 
(LISM) bounds the `there-and-then' primordial abundance from below 
((D/H)$_{\rm P} \geq $(D/H)$_{\rm LISM} = (1.5 \pm 0.1)\times 10^{-5}$).  
As may be seen from Figs.~1 and 2, any lower bound to primordial D will 
provide an upper bound to the baryon-to-photon ratio \cite{rafs}.  

Solar system observations of \3he permit an independent, albeit indirect 
determination of the pre-solar system deuterium abundance \cite{gr}.  
This estimate of the Galactic abundance some 4.5~Gyr ago, while somewhat 
higher than the LISM value, has a larger uncertainty (D/H $= (2.1 \pm 0.5)
\times 10^{-5}$~\cite{gg}).  Within the uncertainty it is consistent with 
the LISM value suggesting there has been only modest destruction of deuterium 
in the last 4.5~Gyr.  There is also a recent measurement of deuterium in the 
atmosphere of Jupiter using the Galileo Probe Mass Spectrometer \cite{jup}, 
(D/H $= (2.6 \pm 0.7) \times 10^{-5}$).

To further exploit the solar system and/or LISM deuterium determinations 
to constrain/estimate the primordial abundance would require corrections 
for the Galactic evolution of D.  Although the simplicity of the evolution 
of deuterium (it is only destroyed) suggests that such correction might 
be very nearly independent of the details of specific chemical evolution 
models, large differences remain between different estimates.  It is 
therefore fortunate that data on D/H in high-redshift (nearly primordial), 
low-metallicity (nearly unevolved) Lyman-$\alpha$ absorbers has become 
available in recent years (\cite{quas1,quas2}).  It is expected 
that such systems still retain their original,  primordial deuterium, 
undiluted by the deuterium-depleted debris of any  significant stellar 
evolution.  That's the good news.  The bad news is that, at present, 
D-abundance determinations are claimed for only three such systems, and 
that the abundances inferred for two of them (along with a limit to 
the abundance for a third) appear to be inconsistent with the abundance 
estimated for the remaining one.  Here we have a prime illustration of 
``precise", but possibly inaccurate cosmological data.  Indeed, there 
is a serious obstacle inherent to using absorption spectra to measure 
the deuterium abundance since the isotope-shifted deuterium absorption
lines are indistinguishable from the corresponding lines in suitably 
velocity-shifted ($-81~kms^{-1}$) hydrogen.  Such ``interlopers" may 
have been responsible for some of the early claims \cite{quas1} of a 
``high" deuterium abundance \cite{steig}.  Indeed, an interloper may 
be responsible for the one surviving high-D claim \cite{webb}.  At 
present it seems that only three good candidates for nearly primordial 
deuterium have emerged from ground- and space-based observations.  It
may be premature to constrain cosmology on the basis of such sparse
data.  Nonetheless, the two ``low-D" systems suggest a primordial 
deuterium abundance consistent with estimates of the pre-Galactic 
value inferred from LISM and solar system data ((D/H)$_{\rm P} = 
(3.4 \pm 0.25) \times 10^{-5}$~\cite{bt3}).  To illustrate the
confrontation of cosmological theory with observational data, 
this D-abundance will be adopted in the following.  However, the 
consequences of choosing the ``high-D" abundance ((D/H)$_{\rm P} 
= (20 \pm 5) \times 10^{-5}$~\cite{webb}) will also be discussed.

\subsection{Helium-4}

As the second most abundant nuclide in the Universe (after hydrogen), 
the abundance of $^4$He can be determined to high accuracy at sites 
throughout the Universe.  However, as stars evolve they burn hydrogen 
to helium and the \4he in the debris of stellar evolution contaminates 
any primordial \4he.  Since any attempt to correct for stellar evolution 
will be inherently uncertain, it is sensible to concentrate on the data
from low-metallicity sites.  Extragalactic regions of hot, ionized gas 
(\hii regions) provide such sites, where the helium is revealed via
emission lines formed when singly and doubly ionized helium recombines.  
As with deuterium, current data provide ambiguous estimates of the
primordial helium abundance.  Since the differences ($\Delta$Y = 0.010) 
are larger than the statistical uncertainties ($\la \pm 0.003$), systematic 
errors likely dominate.  Among the currently most likely sources of such
errors are uncertain corrections for collisional excitation in helium,
uncertain corrections for unseen neutral helium and/or hydrogen, and
underlying stellar absorption (leading to an underestimate of the true
strength of the helium emission lines).  In contrast, since the most 
metal-poor of the observed regions have metallicities of order 1/50 -- 
1/30 of solar, the extrapolation from the lowest metallicity regions 
to truly primordial introduces an uncertainty no larger than the 
statistical error.  

Using published data (\cite{p,evan}) for 40 low-metallicity regions, 
one group \cite{osa} finds: Y$_{\rm P} = 0.234 \pm 0.003$.  In contrast, 
from an independent data set of 45 low-metallicity regions another group 
\cite{iz2} infers Y$_{\rm P} = 0.244 \pm 0.002$.  Clearly, these results 
are statistically inconsistent.  It is crucial that high priority be 
assigned to further \hii region observations to estimate/avoid the systematic 
errors.  Until then, since the error budget for Y$_{\rm P}$ is likely 
dominated by systematic rather than statistical uncertainties, in what 
follows, a generous range for \Yp will be adopted: 0.228 $\leq$ Y$_{\rm P} 
\leq 0.248$.

\subsection{Lithium-7}

Cosmologically interesting lithium is observed in the spectra of nearly 
100 very metal-poor halo stars~\cite{sp, rnb}.  These stars are so 
metal-poor they provide a sample of more nearly primordial material than 
anything observed anywhere else in the Universe; the most metal-poor 
among them have heavy element abundances less than one-thousandth of 
the solar metallicity.  However, these halo stars are also the oldest 
objects in the Galaxy and, as such, have had the most time to modify 
their surface abundances.  So, even though any correction for Galactic 
evolution modifying their lithium abundances may be smaller than the 
statistical uncertainties of a given measurement, the systematic 
uncertainty associated with the dilution and/or destruction of surface 
lithium in these very old stars could dominate the error budget.  
There could be additional errors associated with the modeling of the 
surface layers of these cool, low-metallicity, low-mass stars needed 
to derive abundances from absoprtion-line spectra.  It is also possible 
that some of the Li observed in these stars is non-primordial (\eg that 
some of the observed Li may have been produced post-BBN by spallation 
reactions (the breakup of C, N, and O nuclei into nuclei of Li, Be, 
and B) or fusion reactions ($\alpha + \alpha$ to form $^6$Li and $^7$Li) 
in cosmic-ray collisions with gas in the ISM).  In a recent analysis 
\cite{rbofn}, it is argued that as much as $\sim 0.2$~dex of the observed 
lithium abundance, A(Li) $\equiv 12 + $log(Li/H), could be post-primordial 
in origin.


The very large data set of lithium abundances measured in the warmer 
($T > 5800K$), more metal-poor ([Fe/H] $< -1.3$) halo stars define a 
plateau (the``Spite-plateau"~\cite{sp}) in the lithium abundance -- 
metallicity plane.  Depending on the choice of stellar-temperature scale 
and stellar atmosphere model, the abundance level of the plateau is: 
A(Li) = 2.2 $\pm$ 0.1, with very little dispersion in abundances around 
this plateau value.  The small dispersion provides an important constraint 
on models which attempt to connect the present surface lithium abundances 
in these stars to the original lithium abundance in the gas out of which 
these stars were formed some 10 -- 15 Gyr ago.  ``Standard" (\ie non-rotating) 
stellar models predict almost no lithium depletion and, therefore, are 
consistent with no dispersion about the Spite-plateau.  Although early 
work on mixing in models of rotating stars was very uncertain, recently 
stellar models have been constructed which reproduce the angular momentum 
evolution observed for the much younger, low-mass open cluster stars.  
These models have been applied to the study of lithium depletion in main 
sequence halo stars.  A well-defined lithium plateau with modest scatter 
and a small population of ``outliers" (overdepleted stars), consistent 
with the current data, is predicted for depletion factors between 0.2~dex 
and 0.4~dex \cite{pin1}.

To err on the side of caution, a generous range for the plateau abundance, 
2.1 $\leq $~A(Li) $\leq 2.3$, is adopted.  If depletion is absent, this 
range is consistent with the primordial lithium ``valley'' (see Fig. 4). 
For depletion $\geq 0.2$~dex, the consistent primordial lithium abundances 
bifurcate and move up into the ``foothills", although a non-negligible
contribution from post-BBN lithium could move the primordial abundance 
back down again (Fig. 4).

\section{Confrontation Of Theory With Data}

In the context of the ``standard" model (three families of massless, 
or light, two-component neutrinos), the predictions of BBN (SBBN) 
depend on only one free parameter, the nucleon-to-photon ratio $\eta$.  
The key test of the standard, hot, big bang cosmology is to assess 
if there exists a unique value or range of $\eta$ for which the 
predictions of the primordial abundances are consistent with the 
light element abundances inferred from the observational data.  
From a statistical point of view it might be preferrable to perform 
a simultaneous fit of the inferred primordial abundances of D, \3he, 
\4he, and \7li to the SBBN predictions.  In this manner the ``best 
fit" $\eta$, along with its probability distribution may be found, 
and the ``goodness-of-fit" assessed \cite{crisis}.  However, since 
systematic uncertainties most likely dominate observational errors 
at present, the value of this approach is compromised.  An alternate 
approach is adopted here.

As emphasized earlier, deuterium is an ideal baryometer.  As a first
step the primordial abundance of deuterium inferred from observations
at high redshift will be compared with the SBBN prediction to identify
a consistent range for $\eta$.  Then, given this range, the SBBN
abundances of \4he and \7li are predicted and these are compared to 
the corresponding primordial abundances derived from the observational 
data.  The challenge is to see if the D-identified range for $\eta$
leads to consistent predictions for \4he and \7li.  Recall that due to 
its complicated evolutionary history, it is difficult to use \3he to 
test and constrain SBBN.  Furthermore, another consistency test is to 
compare the SBBN-inferred $\eta$ range with the present baryon density
derived from non-BBN observations and theory.  Is our model for the
very early evolution of the Universe consistent with the present
Universe?

From the two well observed, high redshift absorption line systems with
``low-D", the estimate adopted for the primordial-D abundance is: (D/H)
$_{\rm P} = 2.9 - 4.0 \times 10^{-5}$ (see Fig. 3).  Also shown for 
comparison in Figure 3 is the allowed range of the primordial deuterium 
abundance suggested by the ``high-D" abundance inferred from observations 
of one lower redshift absorption-line system.  With allowance for the
$\sim 8\%$ uncertainty in the theoretically predicted abundance, the 
favored range (low-D) for $\eta$ is quite narrow: $\eta_{10} = 5.1 \pm 
0.36$.  It is clear from Figure 4 that for the baryon abundance in this 
range, the BBN-predicted lithium abundance is entirely consistent with 
the Spite-plateau value, even if the plateau were raised by $\sim 0.2$~dex 
to allow for modest stellar destruction/dilution or lowered by a similar 
amount due to post-BBN production.  For this narrow range in $\eta$ the 
predicted \4he mass fraction varies very little.  For $\eta_{10} \approx 
5$, $\Delta$\Yp $\approx 0.010\Delta\eta/\eta$, so that including the 
error in the predicted abundance, \Yp = $0.247 \pm 0.001$.  As may be 
verified from Figure 2, this is within (albeit at the high end of) the 
range allowed by the data from the low metallicity, extragalactic \hii 
regions.  Given the current uncertainties in the primordial abundances,
SBBN is consistent with ``low-D/high-$\eta$".  

The significance of this concordance cannot be underestimated.  A glance 
at Fig. 1 provides a reminder of the enormous ranges for the predicted 
primordial abundances.  That the simplest hot, big bang cosmological 
model can account for (``predict"!) 3 independent abundances (4 with 
\3he; although \3he hasn't been employed in this comparison, its predicted 
abundance is consistent with extant observational data) by adjusting 
only one free parameter ($\eta$) is a striking success.  The theory, 
which is in principle falsifiable, has passed the test.  It needn't 
have.  Indeed, future observational data coupled to better understanding 
of systematic errors may provide new challenges.  For example, if in the
future it should be determined that the primordial helium mass fraction
were lower than \Yp = 0.245, this would be inconsistent (within the
errors) with the ``low-D/high-$\eta$" range derived above.  Similarly,
if the best estimate for the D-determined $\eta$ range changed, the
comparison performed above should be repeated.  With this in mind,
what of the ``high-D/low-$\eta$" range which has been set aside in
the current comparison?

If, in contrast to the deuterium abundance adopted above, the true
value were higher, (D/H)$_{\rm P} = 10 - 30 \times 10^{-5}$, the 
SBBN-favored range in $\eta$ would be lower (see Fig. 2).  Accounting
for the $\sim 8\%$ uncertainty in the theoretically predicted abundance,
$\eta_{10} = 1.7 \pm 0.28$.  Inspection of Figures 2 -- 4 reveals that
as along as $\eta_{10} \ga 1.1 - 1.3$, consistency with \4he and \7li
can be obtained.  Hence, for ``high-D" as well, the standard model 
passes the key cosmological test.  

\subsection{Comparison Of BBN With Non-BBN Baryon Density Estimates}

Having established the internal consistency of primordial nucleosynthesis
in the standard model, it is necessary to proceed to the next key test.
Does the nucleon abundance inferred from processes which occurred during
the first thousand seconds of the evolution of the Universe agree with
estimates/bounds to the nucleon density in the present Universe?

It is a daunting task to attempt to inventory the baryons in the Universe.
Since many (most?) baryons may be ``dark", such approaches can best set
{\it lower} bounds to the present ratio of baryons-to-photons.  One such
estimate \cite{ps} suggests a very weak lower bound on $\eta$ of: $\eta_{10} 
\geq 0.25$, entirely consistent with the BBN estimates above.  Others 
\cite{fhp} have used more subjective (although cautious) estimates of 
the uncertainties, finding a much higher lower bound to the global budget 
of baryons: $\eta_{10} \geq 1.5$, which is still consistent with the 
``low-$\eta$'' range identified using the high-D results.

A possible challenge to the ``low-$\eta$'' case comes from an analysis 
\cite{shf} which employed observational constraints on the Hubble parameter, 
the age of the Universe, the structure-formation ``shape" parameter, and 
the X-ray cluster gas fraction to provide non-BBN constraints on the present 
density of baryons, finding that $\eta_{10} \geq 5$ may be favored over 
$\eta_{10} \leq 2$.  Even so, a significant low-$\eta$, high-D range still 
survives.

\section{Cosmology Constrains Particle Physics}

Limits on particle physics beyond its standard model are mostly 
sensitive to the bounds imposed on the \4he abundance.  As described 
earlier, the \4he abundance is predominantly determined by the 
neutron-to-proton ratio just prior to nucleosynthesis.  This ratio 
is determined by the competition between the weak interaction rates 
and the universal expansion rate.  The latter can be modified from 
its standard model prediction by the presence of ``new" particles 
beyond those known or expected on the basis of the standard model 
of particle physics.  For example, additional neutrino ``flavors" 
($\Delta N_{\nu} > 0$), or other new particles, would increase the 
total energy density of the Universe, thus increasing the expansion 
rate (see equations 4 \& 5), leaving more neutrons to form more \4he.  
For $\Delta N_{\nu}$ sufficiently small, the predicted primordial 
helium abundance scales nearly linearly with $\Delta N_{\nu}$: 
$\Delta$Y $\approx 0.013\Delta N_{\nu}$.  As a result, an {\it upper} 
bound to \Yp coupled with a {\it lower} bound to $\eta$ (since \Yp 
increases with increasing baryon abundance) will lead to an {\it upper} 
bound to $\Delta N_{\nu}$ and a constraint on particle physics~\cite{ssg}.

\begin{figure}[ht]
	\centering
	\epsfysize=4.5truein 
\epsfbox{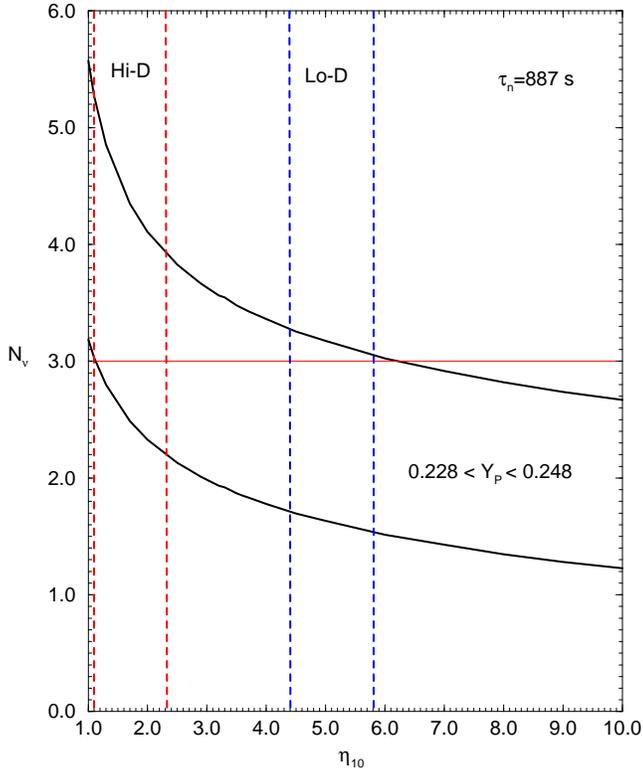}
	\caption{The number of equivalent massless neutrinos $N_{\nu} 
	\equiv 3 + \Delta N_{\nu}$ (see eq.~4) as a function of $\eta$.  
	The region allowed by an assumed primordial \4he mass
	fraction in the range 0.228 -- 0.248 lies between the two solid 
	curves.  The vertical bands bounded by the dashed lines show 
	generous bounds on $\eta$ from ``high" and ``low" deuterium.}
	\label{fig5}
\end{figure}

The constraints on $N_{\nu} = 3 + \Delta N_{\nu}$ as a function 
of the baryon-to-photon ratio $\eta$ is shown in Figure 5.  For 
``low-D/high-$\eta$" there is little room for any ``extra" particles: 
$\Delta N_{\nu} \la 0.3$.  This would eliminate a new neutrino flavor 
($\Delta N_{\nu}  = 1$) or a new scalar particle ($\Delta N_{\nu}  
= 4/7$) provided they were massless or light ($m \ll 1$~MeV), and 
interacted at least as strongly as the ``ordinary" neutrinos.  In
contrast, much weaker constraints are found for ``high-D/low-$\eta$" 
where $N_{\nu}$ as large as 5 ($\Delta N_{\nu} \la 2$) may be allowed 
(see Fig. 5).
  
\section{SUMMARY}

The standard, hot, big bang cosmological model is simple (assuming
isotropy, homogeneity, Newtonian/Einsteinian gravity, standard particle 
physics, etc.) and, likely, simplistic.  In broad brush it offers 
a remarkably successful framework for understanding observations of 
the present and recent Universe.  It may seem hubris to expect that 
this model could provide a realistic description of the Universe during 
the first 20 minutes or so in its evolution when the temperature and 
density were enormously larger than today.  According to the standard 
model, during these first 20 minutes the entire Universe was a nuclear 
reactor, turning neutrons and protons into the light nuclides.  This 
prediction presents the opportunity to use observations here and now 
to test the theory there and then.  As described in this article, SBBN 
predicts observable primordial abundances for just four light nuclides 
D, \3he, \4he, and \7li, as a function of only one adjustable parameter, 
$\eta$, the nucleon-to-photon ratio, which is a measure of the universal 
baryon abundance.  Even though it is currently difficult to use the
extant observational data to bound the primordial abundance of \3he,
SBBN is still overconstrained, yielding three predicted abundances for
one free parameter.  Furthermore, the baryon density inferred from SBBN
must be consistent with that derived from observations of the present
Universe.  Given the many possibilities that SBBN could be falsified 
by the empirical data, it is a remarkable success of the standard model
that there is consistency between theory and observations provided that
there are a few billion photons (most of them in the 2.7~K cosmic 
background radiation) for every neutron or proton (nucleon) in the
Universe.  

This success establishes primordial nucleosynthesis as one of the main
pillars of our standard model of cosmology, providing the only probe of 
the physical Universe during its very early evolution.  Alternate theories
of gravity and/or particle physics must now be tested against the impressive 
success of SBBN.


\begin{thebibliography}{}

\bibitem{ssg} G. Steigman, D.N. Schramm, and J. Gunn, {\it Phys. Lett.} 
{\bf B66} (1977) 202.

\bibitem{els} R. Epstein, J. Lattimer and D.N. Schramm {\it Nature} 
{\bf 263} (1976) 198.

\bibitem{hata} N. Hata, R.J. Scherrer, G. Steigman, D. Thomas, and T.P.
Walker, {\it Ap.J.} {\bf 458} (1996) 637.

\bibitem{rafs} H. Reeves, J. Audouze, W. Fowler, and D.N. Schramm, {\it ApJ}
{\bf 179} (1976) 909.

\bibitem{gr} J. Geiss and H. Reeves, {\it A. A.} {\bf 18} (1972) 126.

\bibitem{gg} J. Geiss and G. Gloeckler, {\it Sp. Sci. Rev.} {\bf 84} 
(1998) 239.

\bibitem{jup} P.R. Mahaffy \etal., {\it Sp. Sci. Rev.} {\bf 84} (1998) 251.

\bibitem{quas1} R.F. Carswell, M. Rauch, R.J. Weymann, A.J. Cooke, and
J.K. Webb, {\it MNRAS} {\bf 268} (1994) L1; A. Songaila, L.L. Cowie,  
C. Hogan, and M. Rugers, {\it Nature} {\bf 368} (1994) 599.

\bibitem{quas2} D. Tytler, X.-M. Fan, and S. Burles, {\it Nature} {\bf 381}
 (1996) 207; S. Burles and D. Tytler, {\it Ap.J.} {\bf 460} (1996) 584; S. 
 Burles and D. Tytler, {\it Ap.J.} {\bf 499} (1998) 699; S. Burles and D. 
 Tytler, {\it Ap.J.} {\bf 507} (1998) 732; S. Burles, D. Kirkman, and D. 
 Tytler, {\it Ap.J.} {\bf 519} (1999) 18.

\bibitem{steig} G. Steigman, {\it MNRAS} {\bf 269} (1994) 53.

\bibitem{webb} J.K. Webb, R.F. Carswell, K.M. Lanzetta, R. Ferlet, M.
Lemoine, A. Vidal-Madjar, and D.V. Bowen, {\it Nature} {\bf 388} (1997)
250; D. Tytler, S. Burles, L. Lu, X. M. Fan, A. Wolfe, and B. D. Savage, 
{\it A.J.} {\bf 117} (1999) 63.

\bibitem{bt3} S. Burles and D. Tytler, Proceedings of the Second Oak 
Ridge Symposium on Atomic and Nuclear Astrophysics (ed. P. Mezzacappa; 
IOP: Bristol) (1998) 113.

\bibitem{osa} K.A. Olive and G. Steigman,  {\it Ap.J. Supp.}
 {\bf 97} (1995) 49.

\bibitem{p} B.E.J. Pagel, E.A. Simonson, R.J. Terlevich and M. Edmunds, 
{\it MNRAS} {\bf 255} (1992) 325.

\bibitem{evan} E. Skillman and R.C. Kennicutt, {\it Ap.J.} {\bf 411} (1993)
655;  E. Skillman, R.J. Terlevich,  R.C. Kennicutt, D.R. Garnett,
and E. Terlevich, {\it Ap.J.} {\bf 431} (1994) 172.

\bibitem{iz2} Y.I. Izotov, and T.X. Thuan,
{\it Ap.J.} {\bf 500} (1998) 188.

\bibitem{sp} F. Spite, and M. Spite,  {\it A.A.} {\bf 115} (1982) 357; 
M. Spite, J.P. Maillard, and F. Spite,  {\it A.A.} {\bf 141} (1984) 56; 
F. Spite, and M. Spite,  {\it A.A.} {\bf 163} (1986) 140;
L.M. Hobbs, and D.K. Duncan,  {\it Ap.J.} {\bf 317} (1987) 796;
R. Rebolo, P. Molaro, J.E. and Beckman, {\it A.A.} {\bf 192} (1988) 192;
M. Spite, F. Spite, R.C. Peterson, and F.H. Chaffee Jr., 
{\it A.A.} {\bf 172} (1987) L9; R. Rebolo, J.E. Beckman, and P. Molaro,
{\it A.A.} {\bf 172} (1987) L17; L.M. Hobbs, and C. Pilachowski, 
{\it Ap.J.} {\bf 326} (1988) L23;
L.M. Hobbs, and J.A. Thorburn, {\it Ap.J.} {\bf 375} (1991) 116;
J.A. Thorburn, {\it Ap.J.} {\bf 399} (1992) L83;
C.A. Pilachowski, C. Sneden,and J. Booth, {\it Ap.J.} {\bf 407}
(1993) 699; 
 L. Hobbs, and J. Thorburn, {\it Ap.J.} {\bf 428} (1994) L25;
 J.A. Thorburn, and T.C. Beers, {\it Ap.J.}
{\bf 404} (1993) L13;
F. Spite, and M. Spite, {\it A.A.} {\bf 279} (1993) L9.
J.E. Norris, S.G. Ryan, and G.S. Stringfellow,
{\it Ap.J.} {\bf 423} (1994) 386.

\bibitem{rnb} S. Ryan, J. Norris, and T. Beers, {\it Ap.J.} 
{\bf 523} (1999) 654.

\bibitem{pin1} M.H. Pinsonneault, T.P. Walker, G. Steigman, and 
V.K. Narayanan, {\it Ap.J} {\bf 527} (1999) 180.

\bibitem{rbofn} S. Ryan, T. Beers, K.A. Olive, B.D. Fields, and 
J. Norris, {\it Ap.J.} {\bf 530} (2000) L57.

\bibitem{crisis} N. Hata, R.J. Scherrer, G. Steigman, D. Thomas, and T.P.
Walker, {\it Ap.J.} {\bf 458} (1996) 637.

\bibitem{ps} M. Persic and P. Salucci, {\it MNRAS} {\bf 258} (1992) 14P.

\bibitem{fhp} M. Fukugita, C.J. Hogan, and P.J.E. Peebles, {\it Ap.J.}
{\bf 503} (1998) 518.

\bibitem{shf} G. Steigman, N. Hata, and J.E. Felten, {\it Ap.J.} {\bf 510}
(1999) 564.

\end{thebibliography}
\end{document}